\documentclass[fleqn]{vch-book}
\usepackage{amsmath}
\usepackage{amssymb}
\usepackage{makeidx}
\usepackage{graphicx}
\usepackage{times}
\usepackage{tabularx}
\usepackage{booktabs}
\usepackage{cite}
\newcommand{\RR}{\mathbb{R}}
\newcommand{\ZZ}{\mathbb{Z}}
\newcommand{\QQ}{\mathbb{Q}}
\newcommand{\iu}{\rm i}
\newcommand{\s}{\scriptscriptstyle <}
\newcommand{\g}{\scriptscriptstyle >}

\makeindex

\begin{document}
\section{Combinatorial problems of (quasi-)crystallography}
\authorafterheading{Michael Baake}
\affil{Institut f\"{u}r Mathematik, Universit\"{a}t Greifswald,\\
Jahnstr. 15a, 17487 Greifswald, Germany}
\authorafterheading{Uwe Grimm}
\affil{Applied Mathematics Department, The Open University,\\
Walton Hall, Milton Keynes MK7 6AA, United Kingdom}

\subsection{Introduction}

The discovery of non-periodic solids has motivated the construction of
numerous examples of aperiodic tiling models, and led to the
systematic theory of cut and project or model sets, as they are now
called, see \cite{Moody,MB2,Martin} for details and further
references. Most combinatorial questions of crystallography, such as
sublattice or shelling structures, have a natural analogue for
aperiodic systems (we will explain the technical terms in more detail
later on). However, the traditional methods of crystallography do not
apply. Fortunately, the most important systems possess a high degree
of symmetry, which manifests itself in an intimate relation to
algebraic number theory \cite{P}. This relationship has been exploited
successfully to tackle combinatorial questions for quasicrystals
\cite{MB1,BM,MW,Ple,Al}.

Interestingly, this approach also simplifies the treatment of
crystals.  A fairly common feature is the use of Dirichlet series
generating functions, which emerges from the observation that many
counting functions, when properly normalised, can be expressed in
terms of multiplicative arithmetic functions, see \cite{Apo} for
background material. This leads to a systematic and unified approach
which will be summarised in this article. Besides explicit results,
the generating functions also allow a precise calculation of
asymptotic properties.

There are various other combinatorial problems, such as the
determination of coordination sequences and coronae, the orbit
structure under inflation, or general complexity considerations.  The
results are usually encapsulated in terms of generating functions,
compare \cite{Wilf} for background material. In this expository
article, we will concentrate on problems that are connected with
Dirichlet series generating functions, and only summarize other
developments.

\begin{vchfigure}
\centerline{\includegraphics[width=\textwidth]{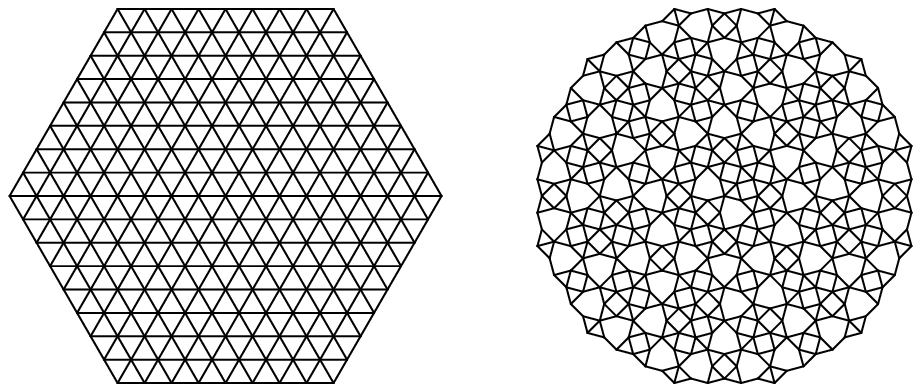}}
\vchcaption{Finite patches of the triangular lattice and 
the twelvefold shield tiling.
\label{tritwelve}\index{shield tiling}\index{lattice!triangular}}
\end{vchfigure}

To be able to explain the concepts in a simple fashion, we shall
mainly use two examples, namely the triangular lattice, written as
$\varGamma=\ZZ[\xi^{}_{3}]$ with $\xi^{}_{3}=\exp(2\pi\iu/3)=
(-1+\iu\sqrt{3}\,)/2$, the set (ring) of Eisenstein integers, and the
vertex set of the twelvefold symmetric shield tiling, see
Fig.~\ref{tritwelve}. We concentrate on explicit results for these
examples, and refer to original sources for a more general exposition
and for details on the asymptotic behaviour.

\subsection{Counting general sublattices}\index{lattice}\index{sublattice}

Let us first consider the triangular lattice
$\varGamma=\ZZ[\xi^{}_{3}]$, or, in fact, any planar lattice, i.e.,
the integer span of two vectors in the plane that are linearly
independent over $\RR$.  Our first question is for the number
$\ell_{2}^{}(m)$ of {\em sublattices\/} $\varGamma'$ of given index
$m=[\varGamma:\varGamma']$. This is a multiplicative function, i.e.,
$\ell_{2}^{}(mn)=\ell_{2}^{}(m)\,\ell_{2}^{}(n)$ for $m$, $n$ coprime.
Due to multiplicativity, it is best encoded in a Dirichlet series
generating function, where one obtains
\cite[Eq.~(3.9)]{MB1}\index{Dirichlet series}
\begin{eqnarray}
  F_{2}^{}(s) &=& \sum_{m=1}^{\infty} \frac{\ell_{2}^{}(m)}{m^s} 
  \;\, = \;\, \zeta(s)\,\zeta(s-1) 
  \;\, = \;\, \prod_{p} \frac{1}{1-p^{-s}}\,\frac{1}{1-p^{1-s}} \\
   &=& {\textstyle 
  1 + \frac{3}{2^s} + \frac{4}{3^s} + \frac{7}{4^s} +  
  \frac{6}{5^s} +  \frac{12}{6^s} +  \frac{8}{7^s} +  \frac{15}{8^s} + 
  \frac{13}{9^s} +  \frac{18}{10^s} + \frac{12}{11^s} + 
  \frac{28}{12^s} 
  + \ldots} \nonumber 
\end{eqnarray} 
Here, $\zeta(s)=\sum_{m=1}^{\infty} m^{-s}$\/ is Riemann's zeta
function, $p$ in the product runs over all rational primes, by which,
following the usual convention, we mean the primes of
$\ZZ$. Furthermore, $\ell_{2}^{}(m)=\sum_{d|m}d$ is the ordinary
divisor function, compare \cite{Apo} for details.\index{zeta
function!Riemann}\index{primes!rational}

The corresponding question is well defined also for
$L=\ZZ[\xi^{}_{12}]$, $\xi^{}_{12}=\exp(2\pi\iu/12)$, when considered
as a (free) $\ZZ$-module of rank $4$.  The number of full rank
submodules of index $m$, $\ell^{}_{4}(m)$, is again multiplicative and
has the generating function \cite[App.~A]{MB1}
\begin{eqnarray}
  F_{4}^{}(s) &=& \sum_{m=1}^{\infty} \frac{\ell_{4}^{}(m)}{m^s} 
  \;\, = \;\, \zeta(s)\,\zeta(s-1)\,\zeta(s-2)\,\zeta(s-3) \\
  &=& {\textstyle 
  1 + \frac{15}{2^s} + \frac{40}{3^s} + \frac{155}{4^s} +  
  \frac{156}{5^s} +  \frac{600}{6^s} +  \frac{400}{7^s} +  
  \frac{1395}{8^s} + \frac{1210}{9^s} +  \frac{2340}{10^s} + 
  \frac{1464}{11^s} 
  + \ldots} \nonumber 
\end{eqnarray} 
where $\ell^{}_{4}(m)=\sum_{d^{}_{1}\cdot\ldots\cdot d^{}_{4}=m}
\,d^{0}_{1}\,d^{1}_{2}\,d^{2}_{3}\,d^{3}_{4}$ with the sum running
over positive $d_{i}$ only. The answer is the same for all
$\ZZ$-modules of rank $4$. They include the cyclotomic rings
$\ZZ[\xi_{n}]$ with $n\in\{5,8,12\}$, which are important for
quasicrystals \cite{BGM}. Note that we follow the mathematical
convention to restrict to $n\not\equiv 2\; (4)$ because
$\ZZ[\xi^{}_{n}] = \ZZ[\xi^{}_{2n}]$ for $n$ odd.

Clearly, this combinatorial question is purely algebraic, and the
method also applies to counting finite index subgroups of finitely
generated free Abelian groups, see \cite[App.~A]{MB1} for details.
The generating function for counting index $m$ subgroups of $\ZZ^n$
reads
\begin{equation}
F_{n}^{}(s) \; = \; \sum_{m=1}^{\infty} \frac{\ell_{n}^{}(m)}{m^s}
\; = \; \zeta(s)\,\zeta(s-1)\cdot\ldots\cdot\zeta(s-n+1)
\end{equation}
and gives the formula $\ell^{}_{n}(m)=\sum_{d^{}_{1}\cdot\ldots\cdot
d^{}_{n}=m} \,d^{0}_{1}\,d^{1}_{2}\cdot\ldots\cdot d^{n-1}_{n}$.

\subsection{Counting similarity sublattices}\index{sublattice!similarity}

Let us turn to (geo)metric properties and ask for the number
$a^{}_{6}(m)$ of {\em triangular\/} sublattices of $\varGamma$ of a
given index $m$, which coincides with the number of ideals of
$\ZZ[\xi^{}_{3}]$ of norm $m$ because $\ZZ[\xi^{}_{3}]$ is a principal
ideal domain \cite{HW}. The answer is given by the Dedekind zeta
function of the cyclotomic (also, quadratic) field
$\QQ(\xi^{}_{3})=\QQ(\sqrt{-3}\,)$, i.e., one obtains
\cite[Eq.~(9)]{MBcol}\index{cyclotomic field}\index{zeta
function!Dedekind}
\begin{eqnarray}
  \zeta^{}_{\QQ(\xi^{}_{3})}(s) &=&
  \sum_{m=1}^{\infty} \frac{a^{}_{6}(m)}{m^s} 
  \;\, = \;\, \frac{1}{1-3^{-s}}\, 
  \prod_{p\equiv 1\,(3)} \frac{1}{\left(1-p^{-s}\right)^{2}}\, 
  \prod_{p\equiv 2\,(3)} \frac{1}{1-p^{-2s}} \\
  &=& {\textstyle 
  1 + \frac{1}{3^s} + \frac{1}{4^s} + \frac{2}{7^s} + 
  \frac{1}{9^s} + \frac{1}{12^s} + \frac{2}{13^s} + \frac{1}{16^s} + 
  \frac{2}{19^s} +  \frac{2}{21^s} + \frac{1}{25^s} 
  + \ldots} \nonumber
\label{sqsim}
\end{eqnarray}
where, e.g., $p\equiv 1(3)$ means that the corresponding product runs
over all rational primes congruent to $1$ modulo $3$.

The corresponding question for the module $\ZZ[\xi^{}_{12}]$ is
answered by the Dedekind zeta function \cite{Wash} of the cyclotomic
field $\QQ(\xi^{}_{12})$, which reads \cite[Eq.~(13)]{MBcol}
\begin{eqnarray}
  \zeta^{}_{\QQ(\xi^{}_{12})}(s) &=&
  \sum_{m=1}^{\infty} \frac{a^{}_{12}(m)}{m^s} \\
  & = & \frac{1}{1-4^{-s}}\, \frac{1}{1-9^{-s}}
  \prod_{p\equiv 1\,(12)} \frac{1}{\left(1-p^{-s}\right)^{4}}\, 
  \prod_{p\equiv -1 \:{\rm or}\: \pm 5\,(12)} \frac{1}{\left(1-p^{-2s}\right)^2}\nonumber \\
  &=& {\textstyle 
  1 + \frac{1}{4^s} + \frac{1}{9^s} + \frac{4}{13^s} + 
  \frac{1}{16^s} + \frac{2}{25^s} + \frac{1}{36^s} + \frac{4}{37^s} + 
  \frac{2}{49^s} +  \frac{4}{52^s} + \frac{4}{61^s}
  + \ldots} \nonumber
\end{eqnarray}

These two generating functions also have an interpretation in terms of
planar colourings \cite{MBcol,Lif,BGS}: $a^{}_{6}(m)$ is (up to
permutation) the number of colourings of the triangular lattice
$\varGamma$ where one colour occupies a similarity (hence triangular)
sublattice of $\varGamma$ of index $m$ and the remaining colours
occupy its cosets. The function $a^{}_{12}(m)$ then counts the
colourings of $\ZZ[\xi^{}_{12}]$ via similarity submodules of index
$m$, all of which are principal ideals of $\ZZ[\xi^{}_{12}]$ as the
latter is a principal ideal domain, see \cite{Wash,BM} for details.
For a discussion of the corresponding colour groups, see \cite{Lif}.

This approach can be extended to all cyclotomic fields with class
number one (the corresponding rings of integers are then principal
ideal domains), see \cite{Wash,MBcol,BG2}, and references given there,
for details.

\subsection{Counting coincidence sublattices}
\index{sublattice!coincidence}

Another geometric problem, with interesting applications to grain
boundaries and twinning phenomena in crystals and quasicrystals, is
the classification of sublattices of $\varGamma$ that can be seen as
the intersection of $\varGamma$ with a rotated copy of itself. These
are the {\em coincidence sublattices\/}, see \cite{MB1} for details,
and the corresponding coincidence index is called $\varSigma$-factor
in materials science.  Once more, the generating function (for the
number of coincidence sublattices of index $m$) is best written as a
Dirichlet series, where, using $\varGamma=\ZZ[\xi^{}_{3}]$, one finds
\begin{eqnarray}
  \Phi^{}_{\ZZ[\xi^{}_{3}]}(s) &=&
  \prod_{p\equiv 1\,(3)} \frac{1+p^{-s}}{1-p^{-s}}
  \;\, = \;\, \frac{1}{1+3^{-s}}\,
  \frac{\zeta^{}_{\QQ(\xi^{}_{3})}(s)}{\zeta(2s)} \\
  &=& {\textstyle 
  1 + \frac{2}{7^s} + \frac{2}{13^s} + \frac{2}{19^s} + 
  \frac{2}{31^s} + \frac{2}{37^s} + \frac{2}{43^s} + \frac{2}{49^s} + 
  \frac{2}{61^s} +  \frac{2}{67^s} + \frac{2}{73^s} 
  + \ldots} \nonumber
\end{eqnarray}
which is derived in \cite[Sec.~IV.A]{Ple}.

Behind this combinatorial problem is a group structure. It turns out
that the set of rotations $R$ such that $\varGamma\cap R\varGamma$ is
a coincidence sublattice of $\varGamma$ forms a group, called
$\mbox{SOC}(\varGamma)$, which is isomorphic to $C^{}_{6}\otimes
\ZZ^{(\aleph^{}_{0})}$, see \cite{BP} for details, in particular on
the structure of this Abelian group. Also, the possible coincidence
indices $[\varGamma : (\varGamma\cap R\varGamma)] =: \varSigma (R)$
(excluding $0$ and $\infty$) form a monoid (a semigroup with unit),
generated by the (rational) primes $p\equiv 1 \; (3)$.

The formulation with cyclotomic integers admits an extension of the
previous results to $\ZZ[\xi^{}_{12}]$, where the generating function
for the number of coincidence submodules of $\ZZ[\xi^{}_{12}]$ of
index $m$ is \cite[Sec.~IV.F]{Ple}
\begin{eqnarray}
  \Phi^{}_{\ZZ[\xi^{}_{12}]}(s) &=&
  \prod_{p\equiv 1\,(12)} \left(\frac{1+p^{-s}}{1-p^{-s}}\right)^{2}
  \,\prod_{p\equiv \pm 5\,(12)} \frac{1+p^{-2s}}{1-p^{-2s}}
  \;\, = \;\, 
  \frac{\zeta^{}_{\QQ(\xi^{}_{12})}(s)}{\zeta^{}_{\QQ(\sqrt{3}\,)}(2s)} \\
  &=& {\textstyle 
  1 + \frac{4}{13^s} + \frac{2}{25^s} + \frac{4}{37^s} + 
  \frac{2}{49^s} + \frac{4}{61^s} + \frac{4}{73^s} + \frac{4}{97^s} + 
  \frac{4}{109^s} +  \frac{4}{157^s} 
  + \ldots} \nonumber
\end{eqnarray}
Here, $\zeta^{}_{\QQ(\sqrt{3}\,)}(s)$ is the Dedekind zeta function of
the quadratic field $\QQ(\sqrt{3}\,)$, i.e.,
\begin{equation}
  \zeta^{}_{\QQ(\sqrt{3}\,)}(s) \; = \; 
  \frac{1}{1-2^{-s}}\, \frac{1}{1-3^{-s}}\,
  \prod_{p\equiv \pm 1\,(12)} \frac{1}{(1-p^{-s})^2}\,
  \prod_{p\equiv \pm 5\,(12)} \frac{1}{1-p^{-2s}}\, .
\end{equation}

There is one subtlety in the application of this result to a {\em
discrete}\/ point set such as the vertex set of the shield tiling.  A
small acceptance correction factor is needed, which can be calculated
explicitly, except for situations where the window has fractal
boundary, as in \cite{BKS}. Further details, together with a general
discussion of the case of $n$-fold symmetry, can be found in
\cite{Ple}.

\subsection{Central shelling}\index{shelling!central}

Let us discuss the shelling problem, first in its version for the {\em
central}\/ shelling. Here, one asks for the number of points of
$\varGamma=\ZZ[\xi^{}_{3}]$ on circles of radius $r$ around the
origin. The result is usually given in terms of lattice theta
functions, see \cite[Ch.~4]{CS} for an extensive exposition. However,
in our situation, it can also be encapsulated in a Dirichlet series.
To this end, one considers only radii $r>0$ and divides the
corresponding shelling number, $c(r^2)$, by $6$, which is the trivial
symmetry factor.  If a shell is non-empty, we have $r^2 = x\,
\overline{x}$ with $x\in\varGamma$, and $r^2=m$ is a (rational)
integer. What then remains is the multiplicative function
$a^{}_{6}(m)$ whose generating function was given above, in a
different context, in Eq.~(\ref{sqsim}).  Further details and
references are given in \cite{BGJR}.

In the case of $\ZZ[\xi^{}_{12}]$, the central shelling function $c$
depends on
$r^2\in\ZZ[\xi^{}_{12}+\overline{\xi}^{}_{12}]=\ZZ[\sqrt{3}\,]$, and
one needs the primes of $\ZZ[\sqrt{3}\,]$ (see \cite{HW} for details)
to derive a formula for $c(r^2)$, compare \cite{BG2}.  Let us write
$\tilde{p}$ for a prime in $\ZZ[\sqrt{3}\,]$, and let $t(\tilde{p})$
be the highest power $t$ such that $\tilde{p}{\,}^t$ divides $r^{2}$,
written as $\tilde{p}{\,}^t \| r^2$.  Then, whenever $t(\tilde{p})$ is
odd for a $\tilde{p}$ that is also a prime in $\ZZ[\xi^{}_{12}]$
(i.e., an inert prime), one has $c(r^2)=0$. Otherwise, one finds the
formula\index{primes!inert}\index{primes!splitting}
\begin{equation} 
   c(r^2) \; = \; 12
   \prod_{\stackrel{\tilde{p}|r^2}
         {\scriptscriptstyle \tilde{p}\; {\rm splits}}}
   \big(t(\tilde{p}) + 1\big)
\label{prime-form}
\end{equation}
where the product runs only over a representative set of those primes
$\tilde{p}$ of $\ZZ[\sqrt{3}\,]$ that split in the extension to
$\ZZ[\xi^{}_{12}]$. In this example, they are precisely the primes
that originate from rational primes $p\equiv 1\; (12)$ and $p\equiv
\pm 5\; (12)$ .

\begin{vchtable}
\renewcommand{\arraystretch}{1.2}
\begin{tabular}{|@{\quad}c@{\qquad}c@{\qquad}c@{\quad}|}
\hline\rule[-1.2ex]{0ex}{4ex}
$\ZZ$ & $\ZZ[\sqrt{3}\,]$ & $\ZZ[\xi^{}_{12}]$ \\
\hline\rule{0ex}{2.8ex}%
$2$ & $2=(2-\sqrt{3}\,)\cdot (1+\sqrt{3})^{2}$ & $1+\sqrt{3}\;$ prime\\
$3$ & $3=(\sqrt{3}\,)^{2}$ & $\sqrt{3}\;$ prime\\
$p\equiv 1\; (12)$ & $p=\tilde{p}\,\tilde{p}'$ & $\tilde{p}=P\,\overline{P}$,
$\;\tilde{p}'=P'\,\overline{P'}$\\
$p\equiv -1\; (12)$ & $p=\tilde{p}\,\tilde{p}'$ & 
$\tilde{p},\tilde{p}'\;$ prime\\
$p\equiv \pm 5\; (12)$ & $p\;$ prime & $p=P\,\overline{P}$%
\rule[-1.2ex]{0ex}{1.2ex} \\
\hline
\end{tabular}
\vchcaption{Splitting structure of primes for $\ZZ[\xi^{}_{12}]$. Here,
${}'$ is a Galois automorphism of $\QQ(\xi^{}_{12})$ which acts as
$\sqrt{3}\mapsto -\sqrt{3}$ in  $\QQ(\sqrt{3}\,)$, and 
$\overline{\protect\phantom{I}}$ denotes complex conjugation. Note that 
$2-\sqrt{3}=(2+\sqrt{3}\,)^{-1}$ is a unit in 
$\ZZ[\sqrt{3}\,]$.\label{splittab}\index{primes!inert}
\index{primes!splitting}}
\renewcommand{\arraystretch}{1}
\end{vchtable}

Fortunately, the function $c(r^2)$ need not be calculated for all
possible values of $r^2$. Since the units of $\ZZ[\sqrt{3}\,]$ are of
the form $\pm (2+\sqrt{3}\,)^{\ell}$ with $\ell\in\ZZ$, all positive
units are actually totally positive, i.e., also their algebraic
conjugate in $\QQ(\sqrt{3}\,)$ is positive. The fundamental unit is
$2+\sqrt{3} = (1+\xi^{}_{12})(1+\overline{\xi}^{}_{12})$, hence all
positive units are of the form $u=x\,\overline{x}$ with
$x\in\ZZ[\xi^{}_{12}]$, and this implies that $c(r^2)$ only depends on
the (principal) ideal of $\ZZ[\sqrt{3}\,]$ that is generated by $r^2$,
compare the proof of theorem~1 in \cite{MW}. So, whenever $s^2 = u
r^2$ with $u$ a positive unit, we have $c(s^2) = c(r^2)$. Also, one
can see that $c\big(r^2\big) = c\big((r^2)'\big)$, where ${}'$ denotes
algebraic conjugation in $\QQ(\sqrt{3}\,)$, defined by the Galois
automorphism $\sqrt{3}\mapsto-\sqrt{3}$, compare Table~\ref{splittab}.

Can the result be given in an even simpler way? Suppose, for a moment,
that the function $c(r^2)$ would only depend on the {\em norm\/} of
$r^2$, i.e., on $m=r^2\,(r^2)'$, where $m$ is a rational integer. This
would mean that $c(r^2)=12f(m)$ with $f(m)$ a multiplicative
arithmetic function, whose Dirichlet series turns out to be
\begin{eqnarray}  \label{centshell}
  C^{}_{12}(s) &=& \sum_{m=1}^{\infty} \frac{f(m)}{m^s} \\[1mm]
  &=& \frac{1}{1-4^{-s}}\,\frac{1}{1-9^{-s}}\,
  \prod_{p\equiv 1\,(12)} 
  \frac{1}{\left(1-p^{-s}\right)^2}\nonumber\\
  &&\times 
  \prod_{p\equiv -1\,(12)}\frac{1}{1-p^{-2s}}\,
  \prod_{p\equiv \pm 5\,(12)}\frac{1}{\left(1-p^{-2s}\right)^2} 
  \nonumber \\[2mm]
  &=& {\textstyle 
  1 + \frac{1}{4^s} + \frac{1}{9^s} + \frac{2}{13^s}  + 
  \frac{1}{16^s} + \frac{2}{25^s} + \frac{1}{36^s} + 
  \frac{2}{37^s} + \frac{2}{49^s} + \frac{2}{52^s} + \frac{2}{61^s}
  + \ldots} \nonumber
\end{eqnarray}
As can be seen from the exact shelling formula (\ref{prime-form}),
this cannot provide the general answer. However, it is correct for
many cases, and is then by far the simplest way to calculate $c(r^2)$.

The first non-trivial value of $m$ where this fails is
$m=169=13^2$. Observing that $13 = (4+\sqrt{3}\,)(4-\sqrt{3}\,) =
\tilde{p}\tilde{p}'$ in $\ZZ[\sqrt{3}\,]$, the possible values of
$r^2$ with norm $m$ are $\tilde{p}^2$, $\tilde{p}\tilde{p}'$ and
$(\tilde{p}')^2$, with shelling numbers 36, 48 and 36, respectively.
Since $f(169)=3$, we get a wrong answer for the middle case this
way. This is an example with a prime $p\equiv 1 \; (12)$. The
contributions to non-empty shells from the other rational primes are
properly counted by the Dirichlet series (\ref{centshell}) because
they either correspond to a unique prime in $\ZZ[\sqrt{3}\,]$ or to
primes which do not split in the step from $\ZZ[\sqrt{3}\,]$ to
$\ZZ[\xi^{}_{12}]$, see Table~\ref{splittab}.
 
So, let us assume that $r$ is the radius of a non-empty shell, i.e.,
$t(\tilde{p})$ is even for all inert primes $\tilde{p}$ of
$\ZZ[\sqrt{3}\,]$.  Then, $12f(m)$ with $m=r^2\,(r^2)'$ gives the
correct shelling number if, whenever $m$ is divisible by a rational
prime $p\equiv 1\; (12)$, $r^2$ is divisible by either $\tilde{p}$ or
$\tilde{p}'$, compare Table~\ref{splittab}, but never by both. Beyond
this situation, one has to use formula (\ref{prime-form}).

Eq.~(\ref{prime-form}), together with the partial simplification of
Eq.~(\ref{centshell}), gives the result for the central shelling of
the full module.  When passing to a model set \cite{Moody}, which is a
discrete subset of the full module, an obvious selection takes place
for the possible radii. Furthermore, a correction factor may become
necessary which depends on the window chosen, see \cite{BG2} for
details. A systematic approach to a complete generating function would
employ summatory functions of arithmetic characters \cite{Wash}, which
is postponed to future work.

\subsection{Averaged shelling}\index{shelling!averaged}

In this last application, we are interested in the {\em average}
number of points on shells of radius $r$, where the average is taken
over all points of our point set as possible centres.  In the lattice
situation, the averaged and central shelling numbers coincide, but
this is not so for model sets \cite{BGJR,Moody}. To be specific, we
explain this for the vertex set of the shield tiling. As for model
sets in general, all averages needed are well defined and unique.

\begin{vchfigure}
\centerline{\includegraphics[width=\textwidth]{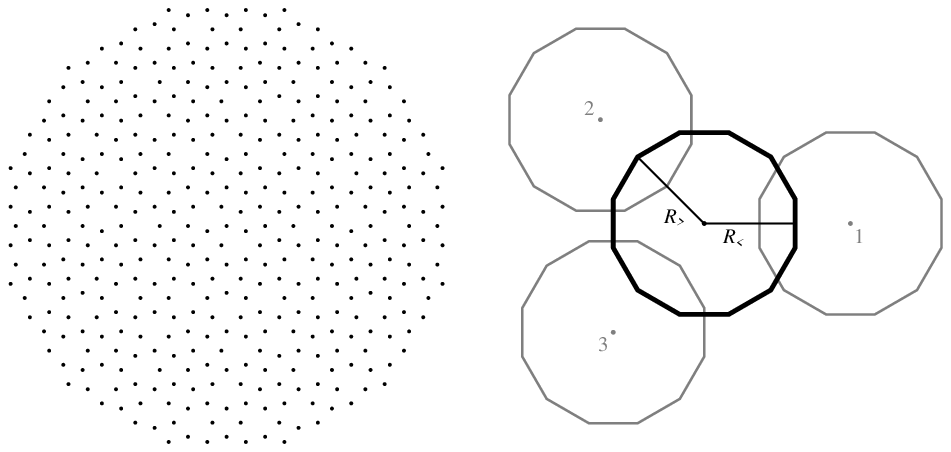}}
\vchcaption{Vertex set of the shield tiling and its dodecagonal 
window (black).\index{window}  
Also shown are three shifted copies of the window (grey), see text for
details, as needed for Table~\ref{tab}. \label{stpic}}
\end{vchfigure}

Let us first describe the vertex set of the shield tiling \cite{Franz}
in algebraic terms. We use $L=\ZZ[\xi^{}_{12}]$ as above, together
with the Galois automorphism $\star$ of $\QQ(\xi^{}_{12})$ given by
$\xi^{}_{12}\mapsto(\xi^{}_{12})^{5}$. This is the star map of the
standard model set construction \cite{Moody}, which gives
\begin{equation}
   \varLambda \; = \; \{ x\in L \mid x^{\star}\in W\}
\end{equation}
where the window $W$ is a relatively compact set with non-empty
interior. For the shield tiling, we choose $W$ as a regular dodecagon
of edge length $1$, hence of inradius $R_{\s}=(2+\sqrt{3}\,)/2$ and
circumradius $R_{\g}=\sqrt{2+\sqrt{3}\,}$, see Fig.~\ref{stpic} for
the correct orientation.\index{model set}\index{shield tiling}

If the window is centred at $0$, the model set is singular, while
generic examples are obtained by shifting the window, i.e.,
$\varLambda^{u} \; = \; \{ x\in L \mid x^{\star}\in W+u\}$, for almost
all $u\in\RR^2$.  The shortest distance between points in a generic
$\varLambda^{u}$ is $(\sqrt{3}-1)/\sqrt{2}$. Joining such points by
edges results in a shield tiling; different generic choices of $u$
result in locally indistinguishable shield tilings.

\begin{vchtable}
\renewcommand{\arraystretch}{1.2}
\begin{tabular}{|cccccc|}
\hline\rule[-1.2ex]{0ex}{4ex}
$r^2$ & representative & orbit length & norm & 
shift type & $a(r^2)$ \\
\hline\rule{0ex}{2.8ex}%
$2-\sqrt{3}$  & $1-\xi$                & $12$ &  $1$ & $2$ & 
  $8-2\sqrt{3}$ \\
$4-2\sqrt{3}$ & $\xi+\overline{\xi}-1$ & $12$ &  $4$ & $1$ & 
  $2$ \\
$6-3\sqrt{3}$ & $1-\xi+\xi^2-\xi^3$    & $12$ &  $9$ & $2$ & 
  $4-2\sqrt{3}$ \\
$1$           & $1$                    & $12$ &  $1$ & $1$ & 
  $8$ \\
$5-2\sqrt{3}$ & $2-\xi$                & $24$ & $13$ & $3$ & 
  $10-4\sqrt{3}$ \\
$2$           & $1+\xi^3$              & $12$ &  $4$ & $2$ & 
  $48-24\sqrt{3}$\\
$4-\sqrt{3}$  & $\xi+\xi^2-2$          & $24$ & $13$ & $3$ & 
  $6$\\
$8-3\sqrt{3}$ & $3\xi-\xi^3-1$         & $24$ & $37$ & $3$ & 
  $-76+44\sqrt{3}$\\
$3$           & $\xi+\overline{\xi}$   & $12$ &  $9$ & $1$ & 
  $-4+\frac{16}{3}\sqrt{3}$\\
$7-2\sqrt{3}$ & $2-2\xi-\xi^2$         & $24$ & $37$ & $3$ & 
  $20-\frac{32}{3}\sqrt{3}$ \\
$2+\sqrt{3}$  & $1+\xi$                & $12$ &  $1$ & $2$ & 
  $48-22\sqrt{3}$\\
$4$           & $2$                    & $12$ & $16$ & $1$ & 
  $2+\frac{4}{3}\sqrt{3}$%
\rule[-1.5ex]{0ex}{1.5ex} \\
\hline
\end{tabular}
\renewcommand{\arraystretch}{1}
\vchcaption{Averaged shelling numbers $a(r^2)$ for the shield tiling 
of edge length $2\sin(\pi/12)=\sqrt{2-\sqrt{3}\,}$,
for all possible distances $0<r\le 2$. Representative elements 
(with $\xi=\xi^{}_{12}$) and orbit lengths refer to the point 
symmetry group $D^{}_{12}$ of the tiling. The norm of $r^2$ is
$m=r^2\,(r^2)'$, and the shift types are explained in
Fig.~\ref{stpic}.\label{tab}} 
\end{vchtable}

Table~\ref{tab} lists all distances $0<r\le 2$ between points of
$\varLambda$ which occur with positive frequency, together with a
representing point in $L$. The corresponding orbit length is given
w.r.t.\ the point symmetry group $D^{}_{12}$. Here, it coincides with
the value of the central shelling number $c(r^2)=12f(m)$, as
calculated by means of Eq.~(\ref{prime-form}) or
Eq.~(\ref{centshell}). For larger values of $r^2$, $c(r^2)$ can
comprise several $D^{}_{12}$-orbits, but only complete ones because
the window is twelvefold and reflection symmetric. For the averaged
shelling, we need to calculate the relative overlap area of the window
$W$ and a shifted copy (the so-called covariogram of
$W$),\index{covariogram} where the shift is the star image of the
representing vector \cite{BG2}. For our examples, all three relative
orientations indicated in Fig.~\ref{stpic} emerge, with squared
midpoint distance $(r^2)'$ in $\QQ(\sqrt{3}\,)$. The resulting
averaged shelling number (last column) is always an element of
$\QQ(\sqrt{3}\,)$. More precisely, all averaged shelling numbers are
in $\frac{1}{3}\ZZ[\sqrt{3}\,]$. This follows from the fact that the
frequency module of the vertex set of the shield tiling is
$\frac{1}{36}\ZZ[\sqrt{3}\,]$. The frequency module is the $\ZZ$-span
of the frequencies of all possible finite patches and can be
determined from the topological invariants of the shield tiling
\cite{Franz2}.

For the overlap scenarios $1$ and $2$ of Fig.~\ref{stpic}, where the
shifts are along the two principal reflection axes of the dodecagon,
the covariogram has a relatively simple form. We find
\begin{equation}
h^{}_{1}(s)
\; =\; 
\begin{cases}
1 - \frac{1}{3}s, & 0\le s<1 \\[1mm]
\frac{15+2\sqrt{3}}{18} - \frac{2\sqrt{3}}{9}s +
  \frac{2\sqrt{3}-3}{18} s^{2}, & 1\le s<2R_{\s}-1\\[1mm]
\frac{5+2\sqrt{3}}{6} - \frac{2}{3}s + \frac{2\sqrt{3}-3}{6} s^{2}, &
2R_{\s}-1\le s<2R_{\s}\\[1mm]
0, & 2R_{\s}\le s
\end{cases}
\end{equation}
and
\begin{equation}
h^{}_{2}(s) \; =\; 
\begin{cases}
1 - \frac{\sqrt{2}(\sqrt{3}-1)}{3} s + \frac{7-4\sqrt{3}}{6} s^2, &
0\le s < R_{\g}\\[1mm]
\frac{5+\sqrt{3}}{6} - \frac{\sqrt{2}}{3} s + \frac{2-\sqrt{3}}{6} s^2, &
R_{\g}\le s < R_{\g}+\sqrt{2}\\[1mm]
\frac{4+2\sqrt{3}}{3} - \frac{\sqrt{2}(1+\sqrt{3})}{3} s + 
      \frac{1}{6}s^2, & R_{\g}+\sqrt{2}\le t < 2R_{\g}\\[1mm]
0, & 2R_{\g}\le s
\end{cases}
\end{equation}
These functions have to be multiplied by the orbit length to give the
averaged shelling number, resp.\ the corresponding contribution to it.

In Table~\ref{tab}, the four examples of type $3$, where the shift is
not along a symmetry direction, were calculated separately (and
exactly). We do not give the more involved general formula for the
covariogram of the regular dodecagon, but mention that
\begin{equation}
h(s) \; = \; \frac{2}{\pi} \arccos\Big(\frac{s}{2R}\Big)
- \frac{s}{\pi R}\sqrt{1-\Big(\frac{s}{2R}\Big)^2}\,,
\end{equation}
with $R=\sqrt{3/\pi}\,R_{\g}$, is a very good approximation. It is
obtained by replacing the dodecagon $W$ with a disk of equal area,
hence of radius $R$. This function is also known as Euclid's hat, see
\cite[p.~100]{Gneiting} and references given there for
details.\index{covariogram}

The averaged shelling in one dimension is rather simple and can be
given in closed form \cite{BGJR}, though no appropriate approach via
generating function has been formulated so far. In two dimensions, the
example of the Ammann-Beenker rhombus tiling with eightfold symmetry
is treated in \cite{BGJR,BG2}, while examples with tenfold symmetry
are shown in \cite{BGJR} (rhombic Penrose tiling) and \cite{BG3}
(T\"ubingen triangle tiling). So, with the example shown here, the
standard symmetries in the plane are covered. Again, the understanding
in terms of generating functions is missing, and in three or more
dimensions, the averaged shelling has not been looked at thoroughly so
far.

\subsection{Other combinatorial questions}

Another combinatorial problem, which can be seen as a variant of the
shelling problem, is the determination of the coordination structure
\cite{CS,BG1,BGRJ}.  This needs a graph structure in addition to the
point set, and is usually formulated for tilings. Once again, one can
ask for the number of points in the $n$th coordination shell (or
corona), or for the analogous averaged quantity. The latter is more
suitable for non-periodic (face to face) tilings. If they are
constructed by a primitive inflation rule, or by the standard cut and
project method (e.g., in its dualization version \cite{KS}), we know
that all averages exist and the problem is well posed.

Let us first consider the situation of a lattice. If we denote the
number of points at graph distance $n$ by $s^{}_{n}$, one uses the
ordinary power series generating function $S(x)=\sum_{n=0}^{\infty}
s^{}_{n}\, x^n$.  For the class of root lattices and root graphs,
these generating functions are rather simple and can be given in
closed form, see \cite{CS2,BG1}. For aperiodic tilings, one faces the
same difficulties as for the averaged shelling discussed above. The
rhombic Penrose tiling and its octagonal sibling, the Ammann-Beenker
tiling, are discussed in \cite{BGRJ}. Once again, the averaged
coordination numbers seem to be ``nice'' numbers. This is due to the
structure of the corresponding frequency modules, and can be
quantified \cite{Franz2}.

Closely related is the more general question for the patch counting
function, which is a direct measure of the local complexity of the
structure. For example, it is well known that Sturmian sequences
possess $n+1$ different words of length $n$. This is the smallest
complexity possible for any non-periodic structure in one dimension.
To what extent this can be generalized to higher dimensions is still
unsettled, see \cite{LP2} for details. In order to come to a natural
analogue to the statement about the Sturmian sequences, one could make
use of the knowledge obtained from the shelling function discussed
above. It provides the possible shell radii and hence the sizes of the
patches to be considered.

The method of iterated inflation is, besides the projection method,
the most frequently used approach to generate aperiodic tilings with
long-range order. Under some mild conditions, the resulting structure
is linearly repetitive, see \cite{LP} for details on this
concept. Inflation also acts naturally on the set of all tilings that
are locally indistinguishable (LI) to the fixed point constructed,
i.e., on the entire LI class. In view of the high degree of (local)
repetitivity, one is also interested in the orbit structure of the LI
class under the inflation $I$. Let $a^{}_{n}$ denote the number of
fixed points of $I^n$, $n\ge 1$, and define
\begin{equation}
\zeta^{}_{\rm LI}(z) \; = \; \exp\left(\,\sum_{m=1}^{\infty}
\frac{a^{}_{m}}{m}\, z^{m}\right)
\end{equation}
so that $\frac{z\zeta^{\prime}_{\rm LI}(z)}{\zeta^{}_{\rm LI}(z)} =
\sum_{m=1}^{\infty} a^{}_{m}\, z^{m}$ is the usual power series
generating function of the sequence $\{a^{}_{m}\mid m\ge 1\}$. The
object $\zeta^{}_{\rm LI}(z)$ is called a dynamical zeta function, see
\cite{R} and references given therein for background
material.\index{zeta function!dynamical}

This function can be calculated systematically, as explained in
\cite{AP}. The cycle structure of inflation can also be extracted,
e.g., by means of the Euler product expansion
\begin{equation}
\frac{1}{\zeta^{}_{\rm LI}(z)} \; =\;
\prod_{n=1}^{\infty} {(1-z^{n})}^{c^{}_{n}}
\end{equation}
where $c^{}_{n}$ is the number of $n$-cycles under inflation. Clearly,
$a_{n}=\sum_{m|n}m\,c_{m}$, and an inversion is possible employing the
M\"{o}bius function, compare \cite{Apo}.

\begin{vchtable}
\begin{tabular}{|c|cc|}
\hline\rule[-1.2ex]{0ex}{4ex}
  & Fibonacci & Penrose  \\
\hline\rule{0ex}{5ex}%
LI class & $\displaystyle \frac{1-z}{1-z-z^2}$   & 
   $\displaystyle \frac{(1-z-z^2)^2 (1+z)}{(1-3z+z^2) 
   (1+z-z^2)^3 (1-z)}$ \\[3ex]
torus    & $\displaystyle \frac{1-z^2}{1-z-z^2}$ & 
   $\displaystyle \frac{(1-z-z^2)^2 (1+z-z^2)^2}{(1-3z+z^2) (1-z)^2 (1+z)^4}$%
\rule[-3.4ex]{0ex}{3.4ex} \\
\hline
\end{tabular}
\vchcaption{Dynamical zeta functions for the inflation action on
the Fibonacci chain and the rhombic Penrose tiling, both on the full
LI classes and on their torus parametrizations.\label{zeta-tab}
\index{zeta function!dynamical}} 
\end{vchtable}

If the tiling admits a construction by the projection method, it also
admits a universal torus parametrization \cite{BHP,Martin2}. The
latter will be one-to-one on generic members of the LI class, but
multiple-to-one on singular members. Consequently, since inflation is
compatible with this reduction step, there is another dynamical zeta
function, this time for the orbit structure on the torus. Both types
of zeta functions are known to be rational. Two examples are
summarized in Table~\ref{zeta-tab}, further examples can be found in
\cite{AP,BHP,HRB}.

\subsection{Open problems}

The above examples demonstrate that a systematic and unified approach
to combinatorial problems of (quasi)crystallography is possible, at
least in the planar case. The situation is more involved in dimensions
$d\ge 3$, where satisfactory results so far exist only for the most
symmetric cases, i.e., those with (hyper)cubic or (hyper)icosahedral
symmetry \cite{MB1,BM,BM2,MW,Al}. Still, even some of these cases
leave room for improvement and simplification, e.g., along the lines
mentioned around Eq.~(\ref{centshell}).

On the other hand, model sets with high symmetry are closely related
to lattices in higher dimensions, where many of these questions are
still open, compare \cite{CS}. One may expect some progress at least
for the class of root lattices, hence also for quasicrystals derived
from them \cite{BJKS}. Further progress is also needed in the
investigation of colour symmetry groups, see \cite{Lif} for a summary
of the present state of affairs.

A big mystery is the meaning of the averaged shelling function. For
the standard tilings with all magic properties (inflation rule,
perfect matching rules, pure point diffraction etc.), the averaged
shelling numbers always seem to be ``nice'' (being algebraic integers
or rationals with bounded denominator), while this is not the case for
model sets with generic windows. This phenomenon points towards
another function defined by these numbers, with various analytic and
topological consequences, but we do not know how to substantiate this
at present.

\subsection*{Acknowledgment}

It is a pleasure to thank Robert V.\ Moody, Peter A.\ B.\ Pleasants
and Alfred Weiss for cooperation and helpful discussions.  We are
grateful to Franz G\"{a}hler and Tilmann Gneiting for providing
relevant material and information.  This work was supported by
Deutsche Forschungsgemeinschaft (Ba 1070/7). Finally, we express our
gratitude to the Erwin Schr\"{o}dinger International Institute for
Mathematical Physics in Vienna for support during a stay in autumn
2002, where this manuscript was completed.


\begin{thebibliography}[1]{99}

\bibitem{Moody}
R.~V.~Moody,
\newblock Model sets:\thinspace A Survey. 
\newblock In: F.\ Axel, F.\ D\'enoyer and J.\ P.\ Gazeau, editors,
\newblock {\em From Quasicrystals to More Complex Systems}.
\newblock EDP Sciences, Les Ulis, and
Springer, Berlin (2000) 145--166;
\newblock math.MG/0002020.

\bibitem{MB2}
M.~Baake,
\newblock A guide to mathematical quasicrystals.
\newblock In:  J.-B.\ Suck,  M.\ Schreiber and P.\
H\"{a}ussler, editors,
\newblock {\em Quasicrystals}. 
\newblock Springer, Berlin (2002) 17--48; 
\newblock math-ph/9901014.

\bibitem{Martin}
M.~Schlottmann,
\newblock Cut-and-project sets in locally compact Abelian groups.
\newblock In:  J.\ Patera, editor,
\newblock {\em Quasi\-crystals and Discrete Geometry}.
\newblock Fields Institute Monographs, vol.~10, 
AMS, Providence, RI (1998) 247--264.

\bibitem{P}
P.~A.~B.~Pleasants,
\newblock Designer quasicrystals:\thinspace Cut-and-project sets 
with pre-assigned properties.
\newblock In:  M.\ Baake and R.\ V.\ Moody, editors,
\newblock {\em Directions in Mathematical Quasicrystals}.
\newblock CRM Monograph Series, vol.~13, AMS, Providence, RI (2000) 95--141.

\bibitem{MB1}
M.~Baake,
\newblock Solution of the coincidence problem in dimensions $d\le 4$.
\newblock In: R.\ V.\ Moody, editor,
\newblock {\em The Mathematics of Long-Range Aperiodic Order}.
\newblock NATO ASI Series C 489, Kluwer, Dordrecht (1997) 9--44.

\bibitem{BM}
M.~Baake and R.~V.~Moody,
\newblock Similarity submodules and semigroups. 
\newblock In: J.~Patera, editor,
\newblock {\em Quasicrystals and Discrete Geometry}. 
\newblock Fields Institute Monographs, vol.~10, 
AMS, Providence, RI (1998) 1--13.

\bibitem{MW}
R.~V.~Moody and A.~Weiss,
\newblock On shelling $E_8$ quasicrystals,
\newblock J.\ Number Th.\ {\bf 47} (1994) 405--412.

\bibitem{Ple}
P.~A.~B.~Pleasants, M.~Baake and J.~Roth,
\newblock Planar coincidences for $N$-fold symmetry,
\newblock J.\ Math.\ Phys.\ {\bf 37} (1996) 1029--1058.

\bibitem{Al}
A.~Weiss,
\newblock On shelling icosahedral quasicrystals.
\newblock In: M.\ Baake and R.\ V.\ Moody, editors,
\newblock {\em Directions in Mathematical Quasicrystals}.
\newblock CRM Monograph Series, vol.~13, AMS, Providence, 
RI (2000) 161--176.

\bibitem{Apo}
T.~M.~Apostol,
\newblock {\em Introduction to Analytic Number Theory}.
\newblock Springer, New York (1976).

\bibitem{Wilf}
H.~Wilf,
\newblock {\em Generatingfunctionology}. 
\newblock 2nd ed., Academic Press, Boston, MA (1994).

\bibitem{BGM}
M.~Baake, U.~Grimm and R.~V.~Moody,
\newblock Die verborgene Ordnung der Quasikristalle,
\newblock Spektrum der Wissenschaft (February 2002) 64--74.

\bibitem{HW}
G.~H.~Hardy and E.~M.~Wright,
\newblock {\em An Introduction to the Theory of Numbers}.
\newblock  5th ed., Clarendon Press, Oxford (1979).

\bibitem{MBcol}
M.~Baake,
\newblock Combinatorial aspects of colour symmetries,
\newblock J.\ Phys.\ A: Math.\ Gen.\ {\bf 30} (1997) 2687--2698;
\newblock mp\_arc/02-323. 

\bibitem{Wash}
L.~C.~Washington,
\newblock {\em Introduction to Cyclotomic Fields}.
\newblock  2nd ed., Springer, New York (1997).

\bibitem{Lif}
R.~Lifshitz,
\newblock Theory of color symmetry for periodic and quasiperiodic crystals,
\newblock Rev.\ Mod.\ Phys.\ {\bf 69} (1997) 1181--1218.

\bibitem{BGS}
M.~Baake, U.~Grimm and M.~Scheffer,
\newblock Colourings of planar quasicrystals,
\newblock J.\ Alloys Comp.\ {\bf 342} (2002) 195--197;
\newblock cond-mat/0110654.

\bibitem{BG2}
M.~Baake and U.~Grimm,
\newblock A note on shelling,
\newblock preprint math.MG/0203025.

\bibitem{BP}
M.~Baake and P.~A.~B.~Pleasants,
\newblock Algebraic solution of the coincidence problem in two
and three dimensions,
\newblock Z.\ Naturf.\ {\bf 50a} (1995) 711--717.

\bibitem{BKS}
M.~Baake, R.~Klitzing and M.~Schlottmann,
\newblock Fractally shaped acceptance domains of quasiperiodic 
square-triangle tilings with dodecagonal symmetry,
\newblock Physica A {\bf 191} (1992) 544--558.

\bibitem{CS}
J.~H.~Conway and N.~J.~A.~Sloane,
\newblock {\em Sphere Packings, Lattices and Groups}.
\newblock 3rd ed., Springer, New York (1999).

\bibitem{BGJR}
M.~Baake, U.~Grimm, D.~Joseph and P.~Repetowicz,
\newblock Averaged shelling for quasicrystals,
\newblock Mat.\ Sci.\ Eng.\ A {\bf 294--296} (2000) 441--445;
\newblock math.MG/9907156.

\bibitem{Franz}
F.~G\"ahler,
\newblock Matching rules for quasicrystals:\thinspace 
The composition-decomposition method,
\newblock J.\ Non-Cryst.\ Solids {\bf 153 \& 154} (1993) 160--164.

\bibitem{Franz2}
F.~G\"ahler,
\newblock private communication (2002).

\bibitem{Gneiting}
T.~Gneiting,
\newblock Radial positive definite functions generated by Euclid's hat,
\newblock J.\ Multivariate Anal.\ {\bf 69} (1999) 88--119.

\bibitem{BG3}
M.~Baake and U.~Grimm,
\newblock Quasicrystalline combinatorics,
\newblock preprint mp\_arc/02-392.

\bibitem{BG1}
M.~Baake and U.~Grimm,
\newblock Coordination sequences for root lattices and related graphs,
\newblock Z.\ Kristallographie {\bf 212} (1997) 253--256;
\newblock cond-mat/9706122.

\bibitem{BGRJ}
M.~Baake, U.~Grimm, P.~Repetowicz and D.~Joseph,
\newblock Coordination Sequences and Critical Points.
\newblock In:  S.~Takeuchi and T.~Fujiwara, editors,
\newblock {\em Proceedings of the 6th International Conference on 
Quasicrystals}.
\newblock World Scientific, Singapore (1998) 124--127;
\newblock cond-mat/9809110.

\bibitem{KS}
P.~Kramer and M.~Schlottmann,
\newblock Dualization of Voronoi domains and klotz construction -- A 
general method for the generation of proper space fillings,
\newblock J.\ Phys.\ A:\ Math.\ Gen.\ {\bf 22} (1989) L1097--L1102. 

\bibitem{CS2}
J.~H.~Conway and N.~J.~A.~Sloane,
\newblock Low-dimensional lattices. VII. Coordination sequences,
\newblock Proc.\ Royal Soc.\ London A {\bf 453} (1997) 2369--2389.

\bibitem{LP2}
J.~C.~Lagarias and P.~A.~B.~Pleasants,
\newblock Local complexity of Delone sets and crystallinity,
\newblock Can.\ Math.\ Bulletin {\bf 45} (2002)  634--652;
\newblock math.MG/0105088.

\bibitem{LP}
J.~C.~Lagarias and P.~A.~B.~Pleasants,
\newblock Repetitive Delone sets and quasicrystals,
\newblock preprint math.DS/9909033.

\bibitem{R}
D.~Ruelle,
\newblock Dynamical zeta functions and transfer operators,
\newblock Notices AMS {\bf 49} (2002) 887--895.

\bibitem{AP}
J.~E.~Anderson and I.~F.~Putnam,
\newblock  Topological invariants for substitution tilings and their 
associated $C^{*}$-algebras,
\newblock Ergod.\ Theory \& Dyn.\ Systems {\bf 18} (1998) 509--537.

\bibitem{BHP}
M.~Baake, J.~Hermisson and P.~A.~B.~Pleasants,
\newblock The torus parametrization of quasiperiodic LI-classes,
\newblock J.\ Phys.\ A:\ Math.\ Gen.\ {\bf 30} (1997) 3029--3056;
\newblock mp\_arc/02-168.

\bibitem{Martin2}
M.~Schlottmann,
\newblock Generalized model sets and dynamical systems.
\newblock In:  M.\ Baake and R.\ V.\ Moody, editors,
\newblock {\em Directions in Mathematical Quasicrystals}.
\newblock CRM Monograph Series, vol.~13, AMS, Providence, RI (2000) 143--159.

\bibitem{HRB}
J.~Hermisson, C.~Richard and M.~Baake,
\newblock A guide to the symmetry structure of quasiperiodic tiling classes,
\newblock J.\ Phys.\ I (France) {\bf 7} (1997) 1003--1018;
\newblock mp\_arc/02-180.

\bibitem{BM2}
M.~Baake and R.~V.~Moody,
\newblock Similarity submodules and root systems in four dimensions,
\newblock Can.\ J.\ Math.\ {\bf 51} (1999) 1258--1276; 
\newblock math.MG/9904028.

\bibitem{BJKS}
M.~Baake, D.~Joseph, P.~Kramer and M.~Schlottmann,
\newblock Root lattices and quasicrystals,
\newblock J.\ Phys.\ A:\ Math.\ Gen.\ {\bf 23} (1990) L1037--L1041;
\newblock cond-mat/0006062.

\end{thebibliography}
\end{document}